\documentclass[conference,9pt]{IEEEtran}

\include{preamble}

\usepackage{comment}

\usepackage{enumitem}
\usepackage{titlesec}
\usepackage{booktabs}
\usepackage{tabularx}
\usepackage{graphicx} 
\usepackage{algorithm}
\usepackage{adjustbox}
\usepackage{algpseudocode}
\usepackage{amsmath, amssymb}
\usepackage{subcaption}
\usepackage{xcolor}
\usepackage{multirow}
\usepackage{array}
\usepackage[colorlinks=true, citecolor=blue, linkcolor=blue, urlcolor=blue]{hyperref}
\usepackage[table]{xcolor}
\usepackage{comment}
\usepackage{xspace}

\usepackage{tikz}
\usetikzlibrary{arrows.meta, shapes.geometric, arrows.meta, positioning, fit}
\usetikzlibrary{positioning, calc}
\tikzstyle{arrow} = [thick,->,>=stealth]
\tikzstyle{block1} = [rectangle, rounded corners, text width=2.7cm, minimum height=1cm, align=center, draw=black, fill=orange!20, font=\scriptsize]
\tikzstyle{block2} = [rectangle, rounded corners, text width=2.7cm, minimum height=1cm, align=center, draw=black, fill=purple!20, font=\scriptsize]
\tikzstyle{block3} = [rectangle, rounded corners, text width=2.7cm, minimum height=1cm, align=center, draw=black, fill=blue!20, font=\scriptsize]

\tikzstyle{blueblock} = [rectangle, rounded corners, text width=2cm, minimum height=1cm, align=center, draw=black, fill=	cyan!15, font=\scriptsize]
\tikzstyle{redblock}  = [rectangle, rounded corners, text width=2cm, minimum height=1cm, align=center, draw=black, fill=yellow!15, font=\scriptsize]
\tikzstyle{whiteblock}  = [rectangle, rounded corners, text width=2cm, minimum height=1cm, align=center, draw=black, fill=white, font=\scriptsize]

\usepackage{listings}
\usepackage{xcolor}
\usepackage[most]{tcolorbox}
\usepackage{caption}
\definecolor{codebg}{rgb}{1.0, 0.96, 0.8}

\lstdefinestyle{verilogstyle}{
  language=Verilog,
  basicstyle=\ttfamily\footnotesize,
  keywordstyle=\color{blue},
  commentstyle=\color{gray},
  morekeywords={logic, input, output, assign, always_ff, posedge, begin, end, endmodule},
  columns=fullflexible,
  keepspaces=true,
  backgroundcolor=\color{codebg},
  frame=none,
  breaklines=true,
  showstringspaces=false
}

\usepackage{xcolor}
\usepackage[a4paper, total={184mm,239mm}]{geometry}
\def\BibTeX{{\rm B\kern-.05em{\sc i\kern-.025em b}\kern-.08em
    T\kern-.1667em\lower.7ex\hbox{E}\kern-.125emX}}

\definecolor{S1}{HTML}{E3F2FD}   
\definecolor{S2}{HTML}{E8F5E9}   
\definecolor{RET}{HTML}{FFF3E0}  
\definecolor{INJ}{HTML}{EDE7F6}  
\definecolor{HEAD}{HTML}{FFFDE7} 
\definecolor{OUT}{HTML}{ECEFF1}  

\newcommand{\frw}{\textsc{PRISM}\xspace}

\begin{document}


\title{\frw : Preference-based Reward and Uncertainty-Integrated Synthesis Minimization for RTL PPA Optimization}

\author{
\IEEEauthorblockN{Armin Abdollahi, Negin Ashrafi, Mehdi Kamal, Massoud Pedram}
\IEEEauthorblockA{University of Southern California\\
Los Angeles, CA, USA\\
\{arminabd, ashrafin, mehdi.kamal, pedram\}@usc.edu}
}

\maketitle


\begin{abstract}
This paper introduces a budget-aware, multi-objective framework (called \frw) for RTL optimization that improves power, performance, and area (PPA) while sharply limiting expensive commercial synthesis tool calls. Starting from a synthesizable baseline and functional testbench, the method iterates through LLM-driven rewrite generation, fast compile/simulation equivalence filtering, low-cost proxy feature extraction, and selective commercial synthesis tool (Synopsys Design Compiler (DC)) evaluation. To prioritize which candidates deserve high-quality synthesis, we train a preference-based reward model using pairwise ranking supervision derived from commercial synthesis tool outcomes, and combine ensemble uncertainty with a diversity-aware acquisition rule to avoid early bias and redundant evaluations. A key advantage is that \frw maintains a best-tradeoff set $G$ of non-dominated designs throughout the run: $G$ preserves the best solutions found so far, is fed back into generation and refinement, and enables consistent, cumulative improvements rather than repeatedly restarting from the baseline. Across RTLLM v2.0 modules, the approach consistently yields post-synthesis gains under fixed commercial synthesis tool call budgets, and ablations confirm that reward-guided selection and candidate dropping are central to sample-efficient progress. \frw efficacy is evaluated across multiple technology nodes (7 nm, 15 nm, and 45 nm) using identical pipelines without retraining.
\end{abstract}

\section{Introduction}

Power, performance, and area (PPA) remain the blunt instruments that decide whether an RTL design is merely working or actually shippable. Even for a fixed micro-architecture, small RTL-level structural choices such as sharing arithmetic, factoring enables, simplifying control, and rewriting compares can shift the downstream synthesis result enough to matter in timing closure and power budgets~\cite{yao2024rtlrewriter, abdollahi2025rocketppa}. Beyond RTL rewriting, recent hardware-design efforts have also explored domain-specific acceleration and optimization, including mixed-signal acceleration for edge inference~\cite{abdollahi2024menage} and hybrid approaches for QUBO solving~\cite{abdollahi2025icd}. In practice, engineers iterate with an RTL synthesis engine and its optimization knobs to converge on acceptable tradeoffs; commercial tools such as Synopsys Design Compiler explicitly target concurrent timing/area/power optimization and are widely used as a reference point~\cite{compiler2016synopsys}. 

The catch is that full-fledged synthesis is expensive as an inner loop. Exploring many rewrite directions by running synthesis on everything quickly collapses under runtime, tool capacity, and iteration latency~\cite{ajayi2019openroad}. This makes naive search over RTL transformations unrealistic. The evaluation oracle is accurate but slow, and the combinatorics of code-level edits explode. As a result, practical flows largely rely on heuristics, human insight, and limited what-if exploration, rather than systematic exploration of the RTL rewrite space~\cite{coward2024rover}. Learned pre-synthesis estimators that predict PPA or timing directly from RTL~\cite{abdollahi2025rocketppa, abdollahi2026timingllm} show that fast, learned models can deliver accurate design-quality feedback without running synthesis. \frw follows the same direction, using cheap proxies and a learned reward model to decide which candidates merit a full synthesis run.

Recent progress in code generation motivates treating RTL optimization as a generate and test problem~\cite{abdollahi2026hdlforge}. Many candidate RTL variants can be proposed and the ones that improve PPA can be retained. But for RTL, test is not only functional equivalence~\cite{mishchenko2006improvements}. The expensive part is the PPA measurement under a fixed library, fixed constraints, and a fixed synthesis script. What is needed is a pipeline that enforces correctness cheaply, prioritizes which candidates are worth paying the synthesis cost for, and still explores diverse rewrite directions so the search does not get trapped in one stylistic neighborhood of the code \cite{lee2023customizable}.

This paper presents a time-efficient, budgeted approach to multi-objective RTL optimization that uses full-fledged commercial synthesis sparingly while still targeting simultaneous improvements in area, power, and delay. Starting from a baseline synthesizable design and a functional testbench, the method searches for functionally equivalent variants that improve post-synthesis PPA and outperform prior state of the art approaches under the same tool, library, and constraint settings.

The core pipeline iterates through four stages of candidate generation, fast filtering, cheap proxy extraction, and selective synthesis evaluation, followed by refinements phase around the strongest candidates. A central component is a learned reward model that predicts which surviving candidates are likely to improve PPA before we spend a DC run~\cite{ouyang2022training}. We train this model using preference-style supervision derived from DC outcomes, rather than unstable direct regression of raw PPA across designs. The training objective is pairwise ranking with a logistic loss, aligning naturally with classic pairwise-comparison models, and avoiding absolute scaling across modules and constraints. To make this practical, we instantiate the reward model via parameter-efficient fine-tuning (LoRA), enabling fast scoring over a large candidate pool~\cite{hu2021lora}. We also incorporate uncertainty estimation through an ensemble of independently trained reward models, using disagreement to moderate early-model bias and improve robustness of candidate selection. Candidate selection for DC explicitly balances predicted reward, uncertainty, and diversity to avoid wasting time on near-duplicates~\cite{lakshminarayanan2017simple}.

Because PPA is inherently multi-objective, the output of the search should not be forced into a single scalar winner. Instead, we maintain a running best-tradeoff set of non-dominated designs so the method naturally returns multiple high-quality variants representing different tradeoffs. This matches standard multiobjective optimization logic. When objectives conflict, the meaningful result is a set of non-dominated solutions rather than a single optimum. We additionally feed representatives back into generation and use them as seeds for a refinement stage, concentrating on small, targeted edits around already-competitive designs rather than repeatedly resetting from.

\section{Related Works}

ChipSeek-R1~\cite{chen2025chipseek} and \cite{zhao2025hardware} rely on on-policy RL updates from synthesis-derived rewards, which are sparse and noisy; stabilizing credit assignment typically requires many DC-evaluated rollouts per update. As the policy shifts, the data distribution changes and forces repeated re-synthesis, so DC labels cannot be amortized well—making the approach sample-hungry and hard to scale to large RTL modules. ROVER~\cite{coward2024rover} and ASPEN~\cite{zhang2025aspen} treat PPA optimization as structured, semantics-preserving rewriting (e-graph/equality-saturation plus guided extraction). Their main limitation is coverage and scalability. Effectiveness depends on the rewrite-rule set and extraction heuristics, e-graphs can blow up on large designs, and accurate PPA ranking still typically falls back to expensive synthesis. VeriOpt~\cite{tasnia2025veriopt} and LLM-VeriPPA~\cite{thorat2025llm} represent LLM-in-the-loop PPA optimization via iterative generate--check--refine, typically enforcing compilability/functional constraints before using synthesis feedback to steer toward better PPA. Their main limitation is speed and evaluation efficiency. Without a principled acquisition rule for which candidates deserve DC, they often produce many near-duplicate candidates and multiple refinement rounds to find improvements which is time consuming and inefficient. MCTS~\cite{delorenzo2024make}, and RTLSquad~\cite{wang2025rtlsquad} wrap the LLM with an explicit search controller such as tree search, evolutionary populations, or multi-agent propose and verify loops to improve exploration beyond greedy decoding. Their shared weakness is DC-bounded scaling. Stronger exploration increases the number of candidates that must be evaluated or compared, so runtime and cost grow quickly on large RTL modules. More broadly, multi-objective optimization has also been explored in other domains through decomposition-based and metaheuristic methods~\cite{AlikhaniKoupaei2025,Heydarpoor2020}.

\noindent{The contributions of our work are:}
\begin{itemize}
  \item We propose a budget-aware LLM-in-the-loop RTL optimization framework that jointly targets power, performance, and area (PPA) while sharply reducing reliance on expensive commercial synthesis calls via staged filtering and selective evaluation.
  \item We introduce a preference-learning reward model trained from pairwise comparisons induced by post-synthesis outcomes, enabling reward-guided ranking of candidate RTL rewrites without requiring dense regression targets.
  \item We develop an uncertainty-aware and diversity-aware acquisition/selection strategy that prioritizes high-utility and non-redundant candidates for synthesis evaluation, improving sample efficiency under fixed synthesis budgets.
  \item We present a persistent best-tradeoff archive $G$ (a non-dominated set) that is maintained across iterations and used for warm-starting generation/refinement; unlike prior approaches that may accept edits that worsen PPA, $G$ ensures the reported best-so-far solutions always improve or, at minimum, maintain PPA relative to previously discovered designs.
  \item We evaluate the approach on RTLLM v2.0 benchmark modules across multiple technology libraries, demonstrating consistent post-synthesis gains under budget-matched settings, and we provide ablations quantifying the effects of reward guidance, diversity selection, and candidate dropping.
\end{itemize}

\IEEEpeerreviewmaketitle

\begin{figure*}[t]
    \centering
    \includegraphics[width=0.8\linewidth]{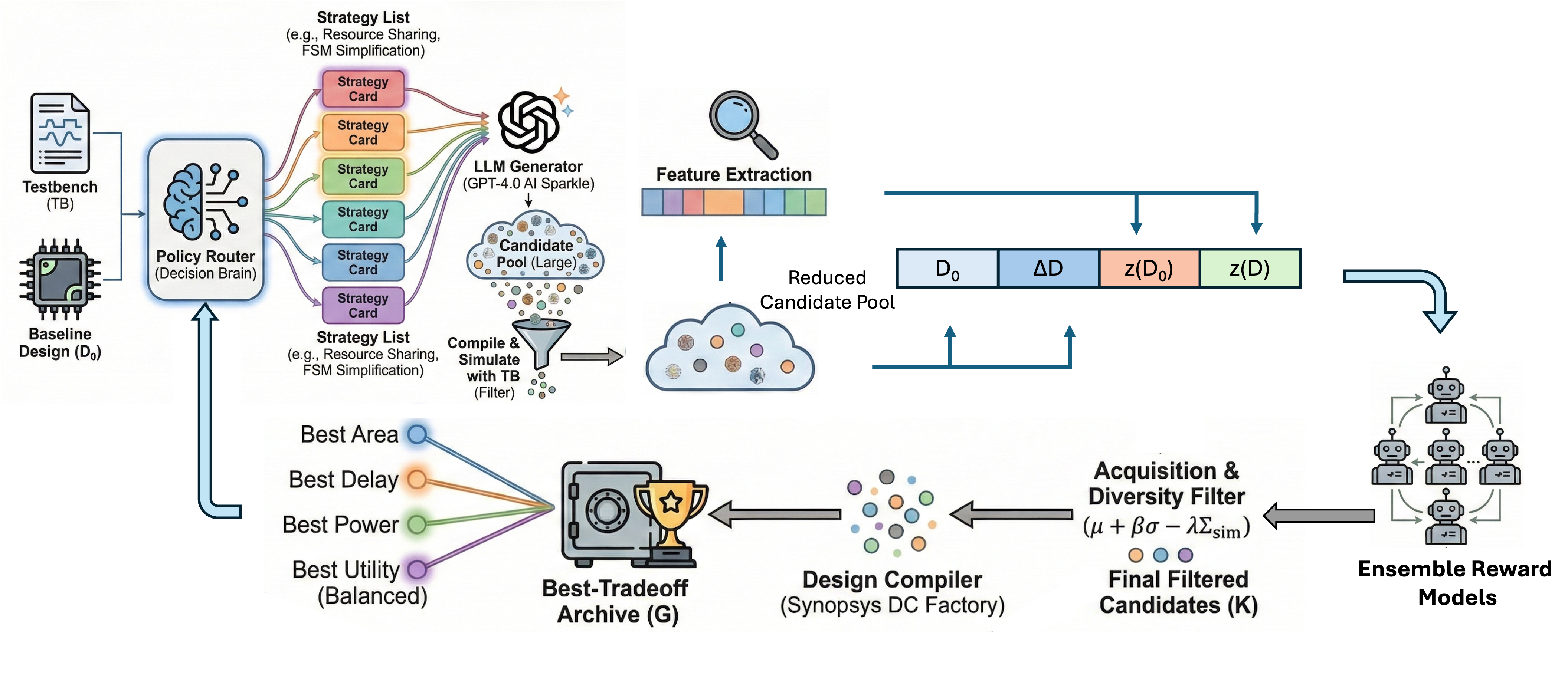}
    \vspace{-20pt}
    \caption{Overview of PPA-Minimizer: closed-loop RTL optimization with LLM candidate generation, reward-model scoring, and DC-in-the-loop evaluation.}
    \label{fig:pdf}
\end{figure*}

\section{Methodology}
\subsection{Problem Formulation: Time-Efficient Multi-Objective RTL Optimization}

Given a baseline synthesizable RTL design $D_0$ and a functional testbench $TB$, \frw's goal is to automatically generate a functionally equivalent RTL variant $D$ by employing the LLMs that improves post-synthesis design quality under standard-cell synthesis. \frw targets multi-objective optimization over area, delay, and power (PPA). The practical challenge is that exploring many rewrite directions by running DC for every candidate is prohibitively slow; therefore, the method is designed to avoid DC evaluation for most candidates by filtering and ranking them using inexpensive checks and learned scoring. Let $E_{DC}(\cdot)$ denote a DC run under a fixed library, constraints, and synthesis script:

\begin{equation}
E_{DC}(D) \to \bigl(A(D),\, P(D),\, Dly(D)\bigr)
\label{eq:edc_metrics}
\end{equation}

\noindent where $A$ is cell area, $P$ is power, and $Dly$ is critical-path delay. Functional correctness is enforced via simulation equivalence on $TB$. Candidates that fail compilation or simulation matching are discarded before DC.

\subsection{Pipeline Overview}
Our proposed method proceeds iteratively through four stages: i) \textbf{Candidate generation}, where \frw produces diverse RTL candidates from a fixed set of rewrite strategies; ii) \textbf{Fast filtering}, where each candidate is compiled and simulated to enforce functional equivalence to $D_0$; iii) \textbf{Cheap hardware proxies and reward scoring}, where proxy features and score candidates are obtained using a fine-tuned reward model; and iv) \textbf{Selective DC evaluation and refinement}, where DC is run on only a small subset of candidates and \frw uses the best-performing results to guide a refinement stage. Throughout, \frw framework maintains a running set of \textbf{best-so-far} designs that capture different PPA tradeoffs (defined precisely in subsection~G); this set is used both to report final results and to guide subsequent generation and refinement. The general flow of the \frw is illustrated in Fig.~\ref{fig:pdf}.

\subsection{Candidate Generation Using Rewrite Prompts}

We define a set of rewrite strategies $S=\{s_1,\ldots,s_{|S|}\}$. Each strategy $s$ is implemented as a rewrite prompt that instructs the generator model to produce a functionally equivalent RTL variant focused on a specific transformation family. At each iteration, we generate candidates by applying a selected list of strategies.
Formally, the candidate pool is:

\begin{equation}
C=\bigcup_{s\in S} \mathrm{Gen}(D_0,G,TB,s;\tau)
\label{eq:candidate_union}
\end{equation}

\noindent where $G$ is the maintained set of best-so-far designs, and $\tau$ is the decoding temperature. We use a strategy set that covers the most common RTL-level sources of unnecessary area/power/delay while remaining compatible with automated equivalence checking via the provided testbench. The strategies are chosen to (i) act on both datapath and control, (ii) encourage synthesis-friendly structure (enables, mux factoring, reduced logic depth), and (iii) provide diverse rewrite modes so the generator does not collapse to near-duplicate edits. These strategies shown in Table~\ref{tab:rewrite_strategies} are applied repeatedly with random seeds and temperatures to produce diversity while staying within synthesizable Verilog constraints. Concretely, each strategy is converted into a natural-language rewrite prompt consisting of the baseline RTL, functional-preservation constraints under the provided testbench, and the strategy-specific instruction summarized in Table~\ref{tab:rewrite_strategies}.

\newcolumntype{Y}{>{\raggedright\arraybackslash}X}
\begin{table}[t]
\centering
\caption{Rewrite strategies used for RTL generation.}
\label{tab:rewrite_strategies}
\scriptsize
\setlength{\tabcolsep}{2pt}
\renewcommand{\arraystretch}{0.9}

\begin{tabularx}{\columnwidth}{@{}p{0.06\columnwidth}p{0.18\columnwidth}YY@{}}
\hline
\textbf{ID} & \textbf{Strategy name} & \textbf{What the prompt instructs} & \textbf{Why it helps PPA} \\
\hline
S1  & Resource sharing & Reuse adders/comparators across mutually exclusive branches; move ops under shared enables & Reduces area/power from duplicated arithmetic; can shorten logic in control-heavy designs \\
S2  & Register merging & Merge redundant state regs; use a single state reg with muxed updates & Cuts FF count and mux overhead; often reduces both area and delay \\
S3  & Enable factoring & Factor common enables/guards; reduce nested if-else logic & Simplifies control cone; can reduce critical-path delay through smaller logic depth \\
S4  & FSM simplification & Merge equivalent states, remove unreachable states, simplify transition conditions & Reduces control logic and improves timing by shrinking next-state logic \\
S5  & Comparator simplification & Replace wide compares with narrower or structured checks (e.g., MSB tests, saturating counters) & Shrinks comparator area and can reduce delay on control paths \\
S6  & Counter optimization & Narrow counter bitwidth, reuse counters, remove redundant counters & Cuts FFs and comparator logic; can reduce both area and delay \\
S7  & Datapath bitwidth narrowing & Reduce internal bitwidths when provably safe under TB behavior & Reduces adder/mux size; often improves delay due to smaller arithmetic \\
S8  & Common subexpression elimination & Hoist repeated expressions into a shared wire/reg & Reduces duplicated logic; improves timing by avoiding repeated cones \\
S9  & Mux factoring / case restructuring & Convert deep if-else chains into case + shared mux patterns & Improves synthesis structure and can reduce delay on select paths \\
S10 & Constant propagation / folding & Precompute constant expressions; remove constant-controlled branches & Removes dead logic and shortens paths; typically helps all PPA axes \\
S11 & Dead logic removal & Remove unused signals/regs and unreachable branches while preserving TB outputs & Cuts area/power; reduces congestion of logic feeding timing paths \\
S12 & Reset/initial cleanup & Simplify reset logic; avoid redundant reset assignments & Reduces control overhead; can reduce delay in reset-driven cones \\
S13 & Pipeline-safe retiming hints & Move registers across purely combinational boundaries without changing I/O behavior under TB & Can reduce critical-path delay when TB tolerance allows (kept conservative) \\
S14 & Sharing of address/control buses & Merge duplicated address/control generation and distribute & Reduces replicated control/data generation logic; can reduce delay in fanout-limited designs \\
S15 & Structured refactor & Re-express logic to expose synthesis optimizations (e.g., balanced trees, associative regrouping) & Helps delay by reducing depth; helps area if it reveals sharing opportunities \\
\hline
\end{tabularx}
\end{table}

\subsection{Fast Functional Filtering and Non-DC Hardware Proxies}
Since the proposed framework does not run DC on all candidates, it first removes incorrect candidates and then obtains proxy features (e.g., \textit{case} statement count) that approximate hardware cost/complexity.

\subsubsection{Functional filtering (compile + simulation equivalence)}
For each candidate $D \in C$, \frw compiles it with a fast simulator (\texttt{iverilog}), simulates under $TB(D)$, and compares normalized output traces against the baseline trace $TB(D_0)$. Candidates that fail compilation, time out, or mismatch the baseline trace are discarded immediately.

\subsubsection{RTL normalization using Yosys}
Many LLM-generated RTL differences are stylistic (temporary wires, redundant assigns, superficial restructuring) and can create noise in proxy extraction and downstream ranking. To reduce this, each candidate is passed through Yosys~\cite{wolf2013yosys} to obtain a more canonical representation. Given RTL text $D$, \frw runs a fixed Yosys script that reads Verilog, resolves hierarchy and sets the top module, converts processes to netlists, applies standard logic optimizations, flattens hierarchy, performs techmapping to Yosys internal primitives, and writes back Verilog without attributes. This produces a normalized RTL that is syntactically cleaner and more consistent across candidates and is used for proxy feature extraction. 

\subsubsection{Hardware-aware proxy features}
For each candidate passing simulation, \frw computes a feature vector $z(D)$ using only non-DC tools.

\paragraph{RTL-structure features (from direct RTL parsing)}
Computed by parsing the Verilog text (token/regex + lightweight syntax scanning): module count, always block count, case statement count; number of blocking/nonblocking assignments; counts of arithmetic ops ($+$, $-$, $*$, shifts), comparisons ($==$, $<$, $>$), bitwise ops; mux-like pattern counts (if-else ladders; ternary operators); inferred register count proxies (nonblocking assignments in clocked always blocks); estimated maximum/minimum bitwidth per key signals (from declared widths and slice usage); and fanout proxy from repeated signal usage counts (text-level approximation).

\paragraph{Yosys statistics (from stat on normalized netlist)}
Computed by running Yosys and extracting statistics: number of internal cells by type (\texttt{\$add}, \texttt{\$sub}, \texttt{\$mux}, \texttt{\$dff}, \texttt{\$logic\_not}, etc.); number of wires, number of memories inferred; depth proxies (where available via simple passes; otherwise approximated via logic cone summaries); and normalized operator footprint as a stable representation of datapath/control complexity.

\frw concatenates the RTL-structure features and Yosys statistics into a fixed-length proxy feature vector $z(D)$, using a consistent ordering of components across all designs. Scalar counts (e.g., block counts and operator counts) are included directly, while Yosys cell-type counts are represented as a vector indexed by the set of supported internal cell types (missing types are assigned zero). The resulting $z(D)$ provides a deterministic, hardware-aware summary of each candidate that can be used for scoring and for computing similarity between candidates.
These proxy features are deterministic outputs of RTL parsing and Yosys analysis. Pairing compact, domain-informed feature sets with learned predictors is a common strategy when inputs are noisy and heterogeneous~\cite{shumilov2025data,jin2026novel}.

\subsection{Reward Model for Multi-Objective Ranking (Area, Power, Delay)}

A learned reward model $R_\phi(\cdot)$ predicts candidate quality by using candidate RTL and proxy features before calling DC. The purpose is to select a small subset of candidates that are most likely to improve PPA.

\subsubsection{Utility definition for training labels (using DC metrics only as supervision)}

Directly regressing raw $A, P, Dly$ across different designs is unstable due to scale variation. Instead, we define a normalized utility relative to each design’s baseline $D_0$. Let $(A_0,P_0,Dly_0)$ be baseline DC metrics and $(A,P,Dly)$ be candidate DC metrics. We define:

\begin{equation}
\begin{aligned}
U(D)=\, & \frac{A_0-A}{A_0}
\;+\;
\frac{P_0-P}{P_0} \\
& \;+\;
\frac{Dly_0-Dly}{Dly_0}
\;.
\end{aligned}
\label{eq:utility_ud}
\end{equation}

This utility is used only to convert DC outcomes into preference labels for training; it is not used as an input to the reward model.

\subsubsection{Preference-based training objective}

We train $R_\phi$ as a ranking model. From candidates evaluated by DC for the same design instance, we form pairs $(D^+,D^-)$ such that $U(D^+)>U(D^-)$. The model is trained to satisfy $R_\phi(D^+)>R_\phi(D^-)$ using the logistic ranking loss:

\begin{equation}
L(\phi)=-\mathbb{E}_{(D^+,D^-)}\Big[\log \sigma\big(R_\phi(x^+)-R_\phi(x^-)\big)\Big].
\label{eq:rank_loss}
\end{equation}

Here, the reward model input $x$ includes the baseline RTL $D_0$ and a compact edit representation $\Delta D$ that explains the candidate RTL $D$ as a difference against $D_0$ to reduce length and emphasize edits, along with structured proxy features $z(D_0)$ and $z(D)$. The $x$, is then passed to the reward head to predict score.

\begin{equation}
x = \big[ D_0 \; ; \; \Delta D \; ; \; z(D_0) \; ; \; z(D) \big]
\label{eq:reward_input}
\end{equation}

\subsubsection{Fine-tuned 7B reward model implementation}

We instantiate $R_\phi$ as a parameter-efficient fine-tuning (LoRA) of Qwen2.5-7B code-capable language model. The model reads the structured input $x$ and outputs a scalar reward score. A learned linear regressor is applied to the final hidden state of the input sequence to obtain the scalar reward. Training uses the pairwise ranking loss above. At inference time, scoring is fast enough to rank a large candidate pool. Steering a model's internal representations instead of updating its weights has proven effective for pre-synthesis timing prediction from Verilog~\cite{abdollahi2026timingllm}, and more recent work steers multiple layers jointly through sparse-autoencoder circuits~\cite{saadatinia2026circuitsteer}; for its reward model, \frw relies on LoRA fine-tuning with a learned scalar head, which provides the fast, stable ranking signal the acquisition step needs.

\subsubsection{Uncertainty Estimation}

Reward scores can be unreliable early in the run or when candidates differ substantially from training patterns. To avoid repeatedly selecting candidates that the model is confidently wrong about, we estimate uncertainty using an ensemble approach. We train $M$ independent reward models $\{R_{\phi_m}\}_{m=1}^M$ using different random seeds (and different LoRA initializations). For each candidate $D$, we compute:

\begin{equation}
\begin{aligned}
\mu(D) &= \frac{1}{M}\sum_{m=1}^M R_{\phi_m}\!\big(x(D)\big), \\
\sigma(D) &= \frac{1}{M-1}\sum_{m=1}^M \Big(R_{\phi_m}\!\big(x(D)\big)-\mu(D)\Big)^2.
\end{aligned}
\label{eq:mu_sigma}
\end{equation}

\noindent $\mu(D)$ captures predicted quality; $\sigma(D)$ captures disagreement across models and is used to occasionally prioritize candidates the model is uncertain about, which improves selection robustness and reduces the chance of getting stuck evaluating only one rewrite style.

\subsection{Selection of Candidates for DC Evaluation}

From the candidate pool $C$ that passes simulation, \frw computes reward scores and selects a small subset for evaluation by DC. The selection criterion balances i) high predicted reward (PPA improvement),
ii) model uncertainty (avoid over-committing to early biases), iii) diversity (avoid sending many near-duplicate edits to DC). For each candidate $D$, an acquisition score is calculated by:

\begin{equation}
a(D)=\mu(D)+\beta\sigma(D).
\label{eq:acq}
\end{equation}

\noindent where $\mu(D)$ and $\sigma(D)$ come from uncertainty estimation (subsection E.4), and $\beta$ is a fixed exploration weight~\cite{auer2002ucb1}. \frw then greedily selects $K$ candidates maximizing total acquisition score while penalizing redundancy using a similarity measure $sim(\cdot,\cdot)$ computed from feature embeddings, where $E$ denotes the set of the $K$ selected candidates:

\begin{equation}
\max_{|E|=K}\;\sum_{D\in E} a(D)\;-\;\lambda\sum_{D_i\ne D_j\in E} sim(D_i,D_j).
\label{eq:select_diverse}
\end{equation}

\begin{equation}
sim(D_i,D_j) \;=\; 
\frac{ z(D_i)^\top z(D_j) }{\|z(D_i)\|\;\|z(D_j)\|}.
\label{eq:sim_cosine}
\end{equation}

This ensures the DC runs cover meaningfully different rewrite directions.

\subsection{Maintaining the Best-Tradeoff Set}

\frw maintains a set $G$, called the best-tradeoff set, containing designs that represent the best PPA tradeoffs observed so far. A design is included in $G$ if it is not strictly worse than another design in all three objectives (area, power, delay). Concretely, for two DC-evaluated designs $D_1$ and $D_2$, we say $D_1$ is strictly worse than $D_2$ if:

\begin{equation}
\begin{aligned}
A(D_1) &\ge A(D_2),\\
P(D_1) &\ge P(D_2),\\
Dly(D_1) &\ge Dly(D_2),\\
\big(A(D_1),P(D_1),Dly(D_1)\big) &\neq \big(A(D_2),P(D_2),Dly(D_2)\big).
\label{eq:ppa_dominance}
\end{aligned}
\end{equation}

In that case, $D_1$ is removed from $G$. Otherwise, it remains as a useful tradeoff point. 
In the proposed framework, we maintain $G$ throughout because the search produces multiple high-quality RTL variants rather than a single clear winner, and $G$ lets the framework report those alternatives without collapsing everything into one arbitrary weighted objective. We also feed representatives from $G$ back into generation so the model can start from strong existing designs instead of repeatedly rewriting from the initial design, which encourages small, cumulative improvements rather than noisy resets. Finally, $G$ provides strong starting points for the refinement in the next iteration, so refinement focuses on polishing already-competitive tradeoffs instead of spending budget rediscovering them.
Finally, we emphasize that the pruning criteria can be relaxed; for example, just power consumption is considered in cases where power consumption is the only design metric of interest.
\subsection{Refinement}

After DC evaluation of the selected set $E$, \frw updates $G$ and then runs a refinement stage to intensify search around the most promising candidates. This stage is designed to reuse the same generation + scoring + selective DC evaluation pattern, but over strong seeds.

\paragraph{Seed selection}
\frw picks a small number of seed designs from the current best-tradeoff set $G$. To ensure coverage of distinct tradeoffs, \frw includes the designs with best area, best power, and best delay in $G$. In addition, it includes the design in $G$ with the highest normalized utility $U(\cdot)$, capturing the strongest balanced improvement observed so far.

\paragraph{Neighborhood generation}
For each seed, \frw generates a neighborhood by re-applying a \emph{subset} of the rewrite strategies $S$, chosen adaptively by the policy router based on the seed’s structure and the current search state. The router allocates a variable number of neighborhood candidates per strategy (including the possibility of allocating none), and when multiple candidates are allocated to the same strategy, \frw diversifies generations by varying decoding temperature. Concretely, we instruct the generator to keep module interfaces unchanged, preserve sequential behavior under $TB$.

\paragraph{Scoring and selection}
All neighborhood candidates pass through the same compile-and-simulate equivalence filter. For survivors, the framework computes proxy features and reward scores. It then selects a small number of diverse neighborhood candidates for DC evaluation.

\paragraph{Acceptance and iteration}
If a neighborhood candidate improves the best-tradeoff set $G$ (i.e., it is not strictly worse than existing members), it is added and may replace dominated designs. The refinement repeats for a small number of steps, enabling a controlled tightening process around good seeds without evaluating every candidate in DC.
This refinement stage is the mechanism that converts coarse, strategy-driven rewrites into consistent improvements by repeatedly proposing and validating small structural changes around already-strong designs.

\subsection{Learned Policy Router}
Instead of allocating candidate generations for each rewrite strategy, we learn a router $\pi_\psi(s\mid D)$ that outputs a distribution over strategies $s\in S$ conditioned only on the current seed RTL module. We train $\pi_\psi$ offline from logged runs by using the observed post-DC utility $U(\cdot)$ as supervision.

For each seed design $D$ and strategy $s$, we define a strategy payoff as the best utility achieved among candidates generated by applying $s$ to $D$:
\begin{equation}
r(D,s) \;=\; \max_{D' \in \mathcal{C}(D,s)} U(D'),
\label{eq:router_payoff}
\end{equation}
where $\mathcal{C}(D,s)$ denotes candidates produced by strategy $s$ from $D$ (after passing the compile/simulation filter). We convert these payoffs into a soft target distribution
\begin{equation}
q(s\mid D) \;=\; \frac{\exp(\, r(D,s))}{\sum_{s'\in S}\exp(\, r(D,s'))},
\label{eq:router_target}
\end{equation}
and train the router by minimizing KL divergence to this target, with a regularizer toward a uniform allocation to preserve exploration:
\begin{equation}
\begin{aligned}
\mathcal{L}_{\text{router}}
&=
\mathbb{E}_{D}\!\left[
\mathrm{KL}\!\left(q(\cdot\mid D)\,\|\,\pi_\psi(\cdot\mid D)\right)
\right] \\
&\quad
+\;
\gamma\,\mathbb{E}_{D}\!\left[
\mathrm{KL}\!\left(\pi_\psi(\cdot\mid D)\,\|\,\mathrm{Unif}(S)\right)
\right].
\end{aligned}
\label{eq:router_loss}
\end{equation}

At inference time, $\pi_\psi(\cdot\mid D)$ determines how generation budget is distributed across strategies. We implement $\pi_\psi$ as a lightweight MLP over $z(D)$, followed by a softmax layer that produces the strategy distribution. When multiple samples are allocated to the same strategy, we diversify outputs by varying the decoding temperature.

\section{Results}

\subsection{Experimental Setup}

We evaluated the proposed PPA optimization framework on the RTLLM v2.0~\cite{openllm_rtl_2025} benchmark, which contains a diverse collection of compact RTL modules representing common hardware building blocks. For each benchmark module, PRISM generates candidate RTL rewrites, filters candidates through compilation and testbench-based functional checks, and prioritizes promising implementations for synthesis evaluation using the learned reward model. We compare the post-synthesis area, power, and delay of the optimized implementations against the original benchmark designs under the same synthesis settings and constraints.

All hardware PPA measurements were obtained using Synopsys Design Compiler (DC). We evaluated the proposed method under multiple technology nodes, including NanGate45 \cite{nangate45}, NanGate15 \cite{martins2015nangate15}, and ASAP7 \cite{vashishtha2017asap7}. Importantly, the same optimization pipeline, reward model, and generation strategy are used across all configurations, without any retraining, fine-tuning, or embedding modification. The technology enters the flow solely through the DC configuration.

The proposed optimization pipeline uses two language models with distinct roles. In this work, \frw uses GPT-4.0~\cite{openai_gpt4_2023} as the primary generation model to propose candidate RTL transformations, while the reward model is a separately trained LoRA-fine-tuned Qwen2.5~\cite{qwen2p5_2024} model used to score and guide candidate selection during iterative refinement. This separation allows the generator to explore a broad space of code edits while the reward model provides a stable optimization signal aligned with PPA objectives.

\subsection{Reward-Model Training Data and Validation Date creation}
We trained the reward model using a large-scale, synthesis-grounded preference dataset constructed offline using GPT4.0. Specifically, we curated approximately 11K Verilog modules paired with passing testbenches, and ran each design through a fixed DC flow to obtain reference reports and baseline $(A,P,Dly)$. For each module, we then generated multiple function-preserving RTL variants that induce diverse PPA trade-offs. The functional equivalence was verified by re-running the corresponding testbench for every variant; each variant was synthesized to obtain its DC report, and we formed the proxy-feature vector $z(D)$. We kept 1k of the dataset for our validation and tuning the hyperparameters, and used 10k for fine-tuning our reward model. The reward-model input $x$ concatenates the baseline RTL $D_0$ and candidate RTL $D$ (explained as a diff against $D_0$ to reduce length and emphasize edits), baseline proxy features $z(D_0)$, and proxy features $z(D)$. We instantiated $R_{\phi}$ as a parameter-efficient LoRA fine-tuning of the Qwen2.5-7B code-capable language model, trained with the pairwise ranking objective provided in Eq.~(4) over DC-derived preferences. Note that at inference time, scoring is sufficiently fast to rank large candidate pools and drive acquisition.

\begin{table}[htbp]
\centering
\caption{Hyperparameters (tuned on the validation split).}
\label{tab:hyperparams}
\setlength{\tabcolsep}{5pt}
\renewcommand{\arraystretch}{1.05}
\begin{tabular}{l p{0.62\columnwidth} c}
\hline
\textbf{Param.} & \textbf{Brief description} & \textbf{Value} \\
\hline
$M$ & Reward-model ensemble size & $5$ \\
$K$ & DC evaluations per iteration & $10$ \\
$\beta$ & Exploration weight in Eq.~(6) & $0.5$ \\
$\lambda$ & Diversity penalty weight in Eq.~(7) & $0.2$ \\
$\gamma$ & Router exploration regularization weight & $0.05$ \\
\hline
\end{tabular}
\end{table}

We tuned the hyperparameters provided in Table~\ref{tab:hyperparams} on the held-out 1K-design validation split (kept separate from the 10K designs used to fine-tune the reward model), selecting the configuration that maximizes validation best-so-far normalized utility under a fixed DC budget. We set $M{=}5$ to obtain a stable disagreement estimate $\sigma(D)$ without excessive training overhead, and $K{=}10$ to keep each DC round small while still covering multiple rewrite directions per iteration. We used $\beta{=}0.5$ to ensure uncertainty meaningfully influences acquisition early in the run (mitigating over-commitment to spurious high-reward modes) while preserving exploitation through $\mu(D)$. We set $\lambda{=}0.2$ as a moderate redundancy penalty that empirically reduces duplicate-at-DC behavior without collapsing selection diversity, and used a $\gamma{=}0.05$ as the router exploration weight.

\subsection{Discussion}

\begin{table*}[t]
\centering
\scriptsize
\setlength{\abovecaptionskip}{2pt}
\setlength{\belowcaptionskip}{2pt}
\setlength{\tabcolsep}{2.2pt}
\renewcommand{\arraystretch}{0.90}
\caption{The design parameters of the RTLLM v2.0 baseline modules and those of optimized ones by \frw, both synthesized with Synopsys Design Compiler using the NanGate45 standard-cell library. For each design, critical-path Delay (ns), Area (\(\mu\)m²), and Total Power (W) are reported.}
\label{table:RTLLM2}

\definecolor{neutyellow}{RGB}{255,248,220} 
\newcommand{\ppacell}[2]{\cellcolor{#1}#2}

\resizebox{0.7\textwidth}{!}{
\footnotesize
\begin{tabular}{l|ccc|ccc}
\hline
\multirow{2}{*}{Design} & \multicolumn{3}{c|}{Baseline numbers (Original Benchmark)} & \multicolumn{3}{c}{Ours} \\
 & Delay (ns) & Area ($\mu m^2$) & Power (W) & Delay (ns) & Area ($\mu m^2$) & Power (W) \\
\hline
asyn\_fifo             & 0.72 & 1397.03 & 7.67e-5  & \ppacell{green!70}{0.25} & \ppacell{green!10}{1369.10} & \ppacell{green!20}{7.21e-5} \\
LFSR                  & 0.06 & 23.41   & 1.88e-6  & \ppacell{green!30}{0.05} & \ppacell{green!10}{22.39}   & \ppacell{green!20}{1.72e-6} \\
right\_shifter        & 0.07 & 36.18   & 2.71e-6  & \ppacell{neutyellow}{0.07} & \ppacell{neutyellow}{36.18} & \ppacell{green!20}{2.39e-6} \\
barrel\_shifter       & 0.19 & 31.39   & 2.48e-5  & \ppacell{neutyellow}{0.19} & \ppacell{green!30}{25.65}   & \ppacell{neutyellow}{2.48e-5} \\
LIFObuffer            & 0.39 & 198.17  & 1.63e-5  & \ppacell{green!30}{0.31} & \ppacell{green!30}{159.19}  & \ppacell{green!40}{1.18e-5} \\
RAM                   & 0.38 & 624.04  & 4.36e-5  & \ppacell{green!40}{0.25} & \ppacell{green!40}{401.24}  & \ppacell{green!30}{3.65e-5} \\
alu                   & 1.95 & 1300.21 & 1.10e-3  & \ppacell{green!20}{1.78} & \ppacell{green!20}{1168.30} & \ppacell{green!40}{6.72e-4} \\
pe                    & 2.81 & 2103.53 & 1.80e-4  & \ppacell{green!30}{2.33} & \ppacell{neutyellow}{2103.53} & \ppacell{green!10}{1.74e-4} \\
instr\_reg            & 0.13 & 117.04  & 8.31e-6  & \ppacell{neutyellow}{0.13} & \ppacell{neutyellow}{117.04} & \ppacell{neutyellow}{8.31e-6} \\
signal\_generator     & 0.34 & 61.71   & 5.15e-6  & \ppacell{green!20}{0.29} & \ppacell{green!20}{58.46}   & \ppacell{green!20}{4.48e-6} \\
square\_wave          & 0.54 & 94.16   & 8.09e-6  & \ppacell{neutyellow}{0.54} & \ppacell{neutyellow}{94.16} & \ppacell{neutyellow}{8.09e-6} \\
calendar              & 0.55 & 154.28  & 1.14e-5  & \ppacell{neutyellow}{0.55} & \ppacell{green!20}{141.94}  & \ppacell{green!40}{7.98e-6} \\
parallel2serial       & 0.05 & 46.55   & 3.68e-6  & \ppacell{green!20}{0.05} & \ppacell{green!10}{44.28}   & \ppacell{green!30}{3.12e-6} \\
pulse\_detect         & 0.04 & 14.63   & 1.17e-6  & \ppacell{green!20}{0.04} & \ppacell{green!20}{12.87}   & \ppacell{green!20}{1.01e-6} \\
serial2parallel       & 0.31 & 157.47  & 1.13e-5  & \ppacell{green!30}{0.25} & \ppacell{green!55}{94.48}   & \ppacell{green!40}{7.91e-6} \\
width\_8to16          & 0.18 & 185.93  & 1.31e-5  & \ppacell{green!20}{0.16} & \ppacell{green!30}{156.18}  & \ppacell{green!20}{1.24e-5} \\
traffic\_light        & 0.45 & 129.54  & 1.04e-5  & \ppacell{green!20}{0.41} & \ppacell{green!20}{122.92}  & \ppacell{green!20}{9.04e-6} \\
edge\_detect          & 0.18 & 17.82   & 1.35e-6  & \ppacell{green!20}{0.16} & \ppacell{green!55}{9.29}    & \ppacell{green!70}{4.73e-7} \\
freq\_divbyfrac       & 0.24 & 44.95   & 3.96e-6  & \ppacell{neutyellow}{0.24} & \ppacell{neutyellow}{44.95} & \ppacell{neutyellow}{3.96e-6} \\
freq\_divbyeven       & 0.20 & 19.15   & 1.54e-6  & \ppacell{green!30}{0.16} & \ppacell{green!40}{14.20}   & \ppacell{green!30}{1.21e-6} \\
JC\_counter           & 0.06 & 340.48  & 2.35e-5  & \ppacell{green!30}{0.05} & \ppacell{green!70}{85.12}   & \ppacell{green!70}{4.70e-6} \\
counter\_12           & 0.11 & 28.99   & 2.28e-6  & \ppacell{green!20}{0.10} & \ppacell{green!20}{25.28}   & \ppacell{green!30}{1.89e-6} \\
up\_down\_counter     & 1.00 & 177.42  & 1.28e-5  & \ppacell{green!30}{0.84} & \ppacell{green!30}{139.22}  & \ppacell{green!55}{7.24e-6} \\
adder\_bcd            & 0.51 & 27.93   & 3.65e-5  & \ppacell{green!20}{0.46} & \ppacell{green!30}{22.82}   & \ppacell{green!30}{2.78e-5} \\
adder\_pipe\_64bit    & 1.09 & 2323.78 & 1.64e-4  & \ppacell{green!20}{0.96} & \ppacell{green!70}{929.51}  & \ppacell{green!20}{1.43e-4} \\
adder\_32bit          & 1.52 & 170.24  & 1.72e-4  & \ppacell{green!40}{1.06} & \ppacell{green!30}{132.79}  & \ppacell{green!30}{1.45e-4} \\
adder\_16bit          & 0.98 & 68.10   & 5.65e-5  & \ppacell{green!55}{0.57} & \ppacell{green!55}{32.91}   & \ppacell{green!40}{3.83e-5} \\
adder\_8bit           & 0.50 & 34.05   & 2.75e-5  & \ppacell{green!55}{0.30} & \ppacell{green!40}{22.28}   & \ppacell{green!40}{1.76e-5} \\
fixed\_point\_adder   & 2.92 & 546.63  & 7.11e-4  & \ppacell{green!70}{0.48} & \ppacell{green!70}{37.89}   & \ppacell{green!70}{3.20e-5} \\
fixed\_point\_subtractor & 2.37 & 416.56 & 4.80e-4 & \ppacell{green!70}{0.38} & \ppacell{green!70}{17.47}   & \ppacell{green!70}{1.10e-5} \\
multi\_pipe\_4bit     & 0.32 & 163.06  & 1.19e-5  & \ppacell{green!40}{0.23} & \ppacell{green!30}{127.19}  & \ppacell{green!30}{9.50e-6} \\
multi\_pipe\_8bit     & 0.86 & 816.62  & 5.51e-5  & \ppacell{green!40}{0.58} & \ppacell{green!40}{582.21}  & \ppacell{green!40}{3.68e-5} \\
multi\_16bit          & 2.85 & 790.02  & 5.98e-5  & \ppacell{green!40}{1.84} & \ppacell{green!20}{687.32}  & \ppacell{green!40}{4.34e-5} \\
comparator\_4bit      & 0.19 & 18.89   & 1.42e-5  & \ppacell{green!30}{0.16} & \ppacell{green!30}{15.39}   & \ppacell{green!40}{9.33e-6} \\
comparator\_3bit      & 0.12 & 13.03   & 1.00e-5  & \ppacell{neutyellow}{0.12} & \ppacell{green!20}{11.39}  & \ppacell{green!20}{9.48e-6} \\
radix2\_div           & 0.70 & 367.08  & 2.92e-5  & \ppacell{green!30}{0.58} & \ppacell{green!30}{289.30}  & \ppacell{green!40}{2.07e-5} \\
div\_16bit            & 7.88 & 898.02  & 2.29e-3  & \ppacell{green!40}{5.62} & \ppacell{green!40}{652.33}  & \ppacell{green!30}{1.83e-3} \\
accu                  & 0.64 & 138.85  & 1.15e-5  & \ppacell{green!20}{0.56} & \ppacell{green!20}{122.91}  & \ppacell{green!20}{1.03e-5} \\
sub\_64bit            & 3.86 & 339.42  & 3.35e-4  & \ppacell{green!30}{2.98} & \ppacell{green!30}{256.87}  & \ppacell{green!40}{2.46e-4} \\
\hline
\end{tabular}
}
\end{table*}

Table~\ref{table:RTLLM2} summarizes post-synthesis PPA for the RTLLM v2.0 benchmark designs, reporting each module’s baseline (Original Benchmark) timing, area, and power alongside optimized implementation under the same evaluation setup. Overall, the optimized variants consistently preserve or improve critical-path delay while also reducing silicon cost and/or power on many designs, for example, the delay of asyn\_fifo is improved from 0.72→0.25 ns with lower area and power, delay of RAM is improved 0.38→0.25 ns with a sizable area reduction, and design parameters of the ALU show simultaneous gains in delay (1.95→1.78 ns), area, and power (1.1e-3→ 6.72e-4 W). Several control-oriented blocks also see meaningful design parameter improvements (e.g., in the cases of the LIFObuffer and serial2parallel, both area and power are improved), while a small subset remains unchanged (e.g., instr\_reg, square\_wave, freq\_divbyfrac), indicating the method is conservative when no clear PPA improvement is available without risking correctness.

\begin{table}[htbp]
\centering
\caption{Average post-synthesis PPA improvements across technology nodes under a budget-matched setting across RTLLM v2.0. Results report mean percentage reductions (↓) in area, total power, and critical-path delay relative to each design's baseline within the same DC flow and constraints for that library.}
\label{tab:node_summary}
\small
\setlength{\tabcolsep}{6pt}
\renewcommand{\arraystretch}{1.1}
\begin{tabular}{lccc}
\hline
\textbf{Configuration} & \textbf{Avg Area} $\downarrow$ & \textbf{Avg Power} $\downarrow$ & \textbf{Avg Delay} $\downarrow$ \\
\hline
NanGate45 (45\,nm) & 23.56\% & 25.12\% & 20.55\% \\
NanGate15 (15\,nm) & 22.61\% & 24.60\% & 21.19\% \\
ASAP7 (7\,nm)      & 22.79\% & 26.01\% & 20.46\% \\
\hline
\end{tabular}
\end{table}

In Table~\ref{tab:node_summary}, across all three technology nodes, the proposed pipeline achieves consistent post-synthesis gains in area, power, and delay across RTLLM v2.0. The improvements remain stable when switching libraries, indicating the method is largely technology-agnostic in practice: the same generation, filtering, reward-guided selection, and refinement loop transfers across nodes.

\begin{figure}[t]
    \centering
    \includegraphics[width=0.9\linewidth]{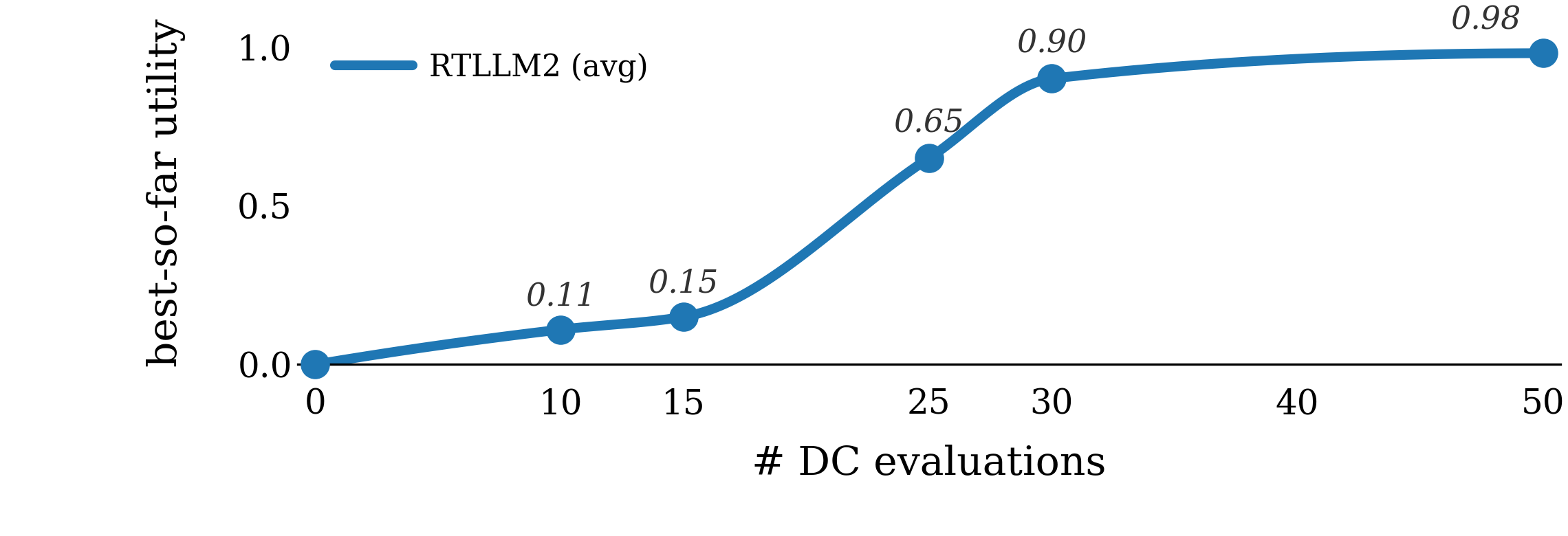}
    \caption{Normalized best-so-far utility vs. DC evaluations.}
    \label{fig:pic1}
\end{figure}

Figure~\ref{fig:pic1} reports DC-budget convergence by plotting the normalized best-so-far utility (baseline mapped to 0) as a function of the number of Design Compiler (DC) evaluations, averaged across RTLLM2 modules. In the early budget ($\approx$0–15 DC calls), improvements are modest because the search is primarily exploring and screening multiple candidate strategies; once promising directions are identified, utility rises sharply ($\approx$15–30 calls) as the method exploits higher-quality transformations, and then gradually saturates as additional DC evaluations yield diminishing returns.

\begin{figure}[t]
    \centering
    \includegraphics[width=0.9\linewidth]{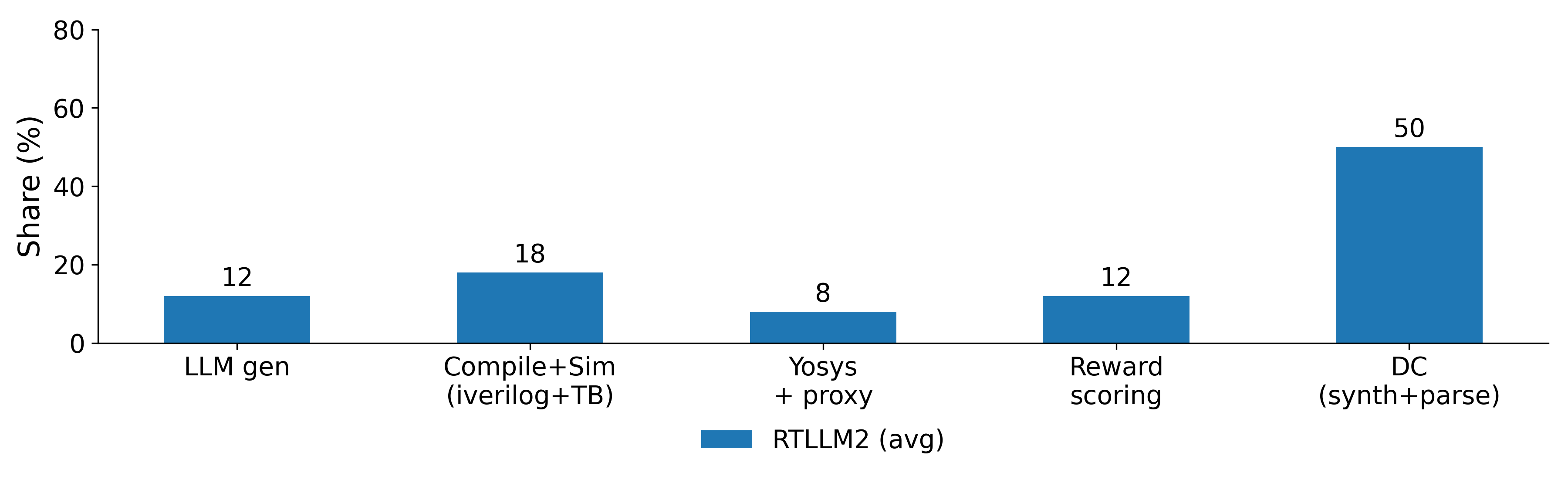}
    \caption{Runtime share breakdown across pipeline stages for RTLLM2 modules}
    \label{fig:pic2}
\end{figure}

Figure~\ref{fig:pic2} decomposes the end-to-end pipeline into five stages—LLM generation, compile/simulation filtering, Yosys-based normalization and proxy extraction, reward scoring, and DC synthesis/report parsing—and reports each stage’s share of total runtime across RTLLM v2.0 benchmark modules. DC synthesis and report parsing account for 50\% of the total runtime, while compile/simulation filtering accounts for 18\%, LLM generation and reward scoring each account for 12\%, and Yosys-based normalization and proxy extraction account for 8\%. Reporting percentages rather than absolute times avoids machine- and license-dependent variability while illustrating the relative computational costs of the pipeline.

We evaluate a set of \emph{budget-matched} baselines under the \emph{same} DC flow, constraints, and RTLLM v2.0 designs, and report average percent improvement over the published baseline PPA. Specifically, \textbf{Random Search} samples candidates uniformly at random and returns the best design observed under the same DC-call budget; \textbf{Generator-only + DC} uses the same LLM candidate generator but selects candidates without a learned reward model (self-scored ranking) and evaluates them in DC; \textbf{Bandit/BO (UCB/TS)} treats a discrete set of rewrite templates/operators as arms and uses DC utility feedback to adaptively allocate evaluation budget via standard bandit/Bayesian-optimization policies; and \textbf{Evolutionary Search} iteratively mutates RTL using the same edit operators and applies selection based on DC utility to form the next population. This analysis isolates the contribution of reward-model-guided acquisition and diversity-aware selection by comparing against representative, reproducible black-box optimization strategies under identical synthesis conditions.

\begin{figure}[t]
    \centering
    \includegraphics[width=0.9\linewidth]{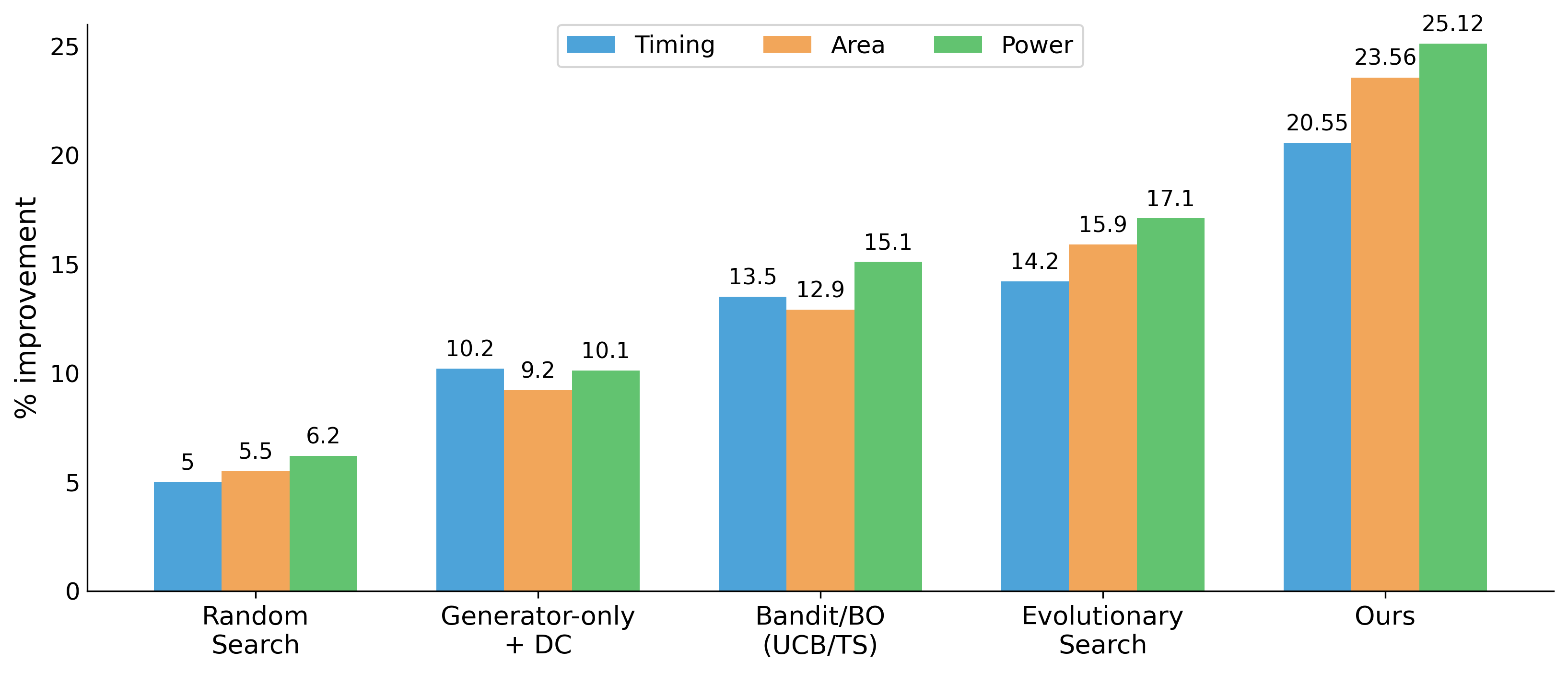}
    \caption{Average \% improvement over RTLLM v2.0 under a budget-matched DC budget.}
    \label{fig:pic3}
\end{figure}

\begin{figure}[t]
    \centering
    \includegraphics[width=0.9\linewidth]{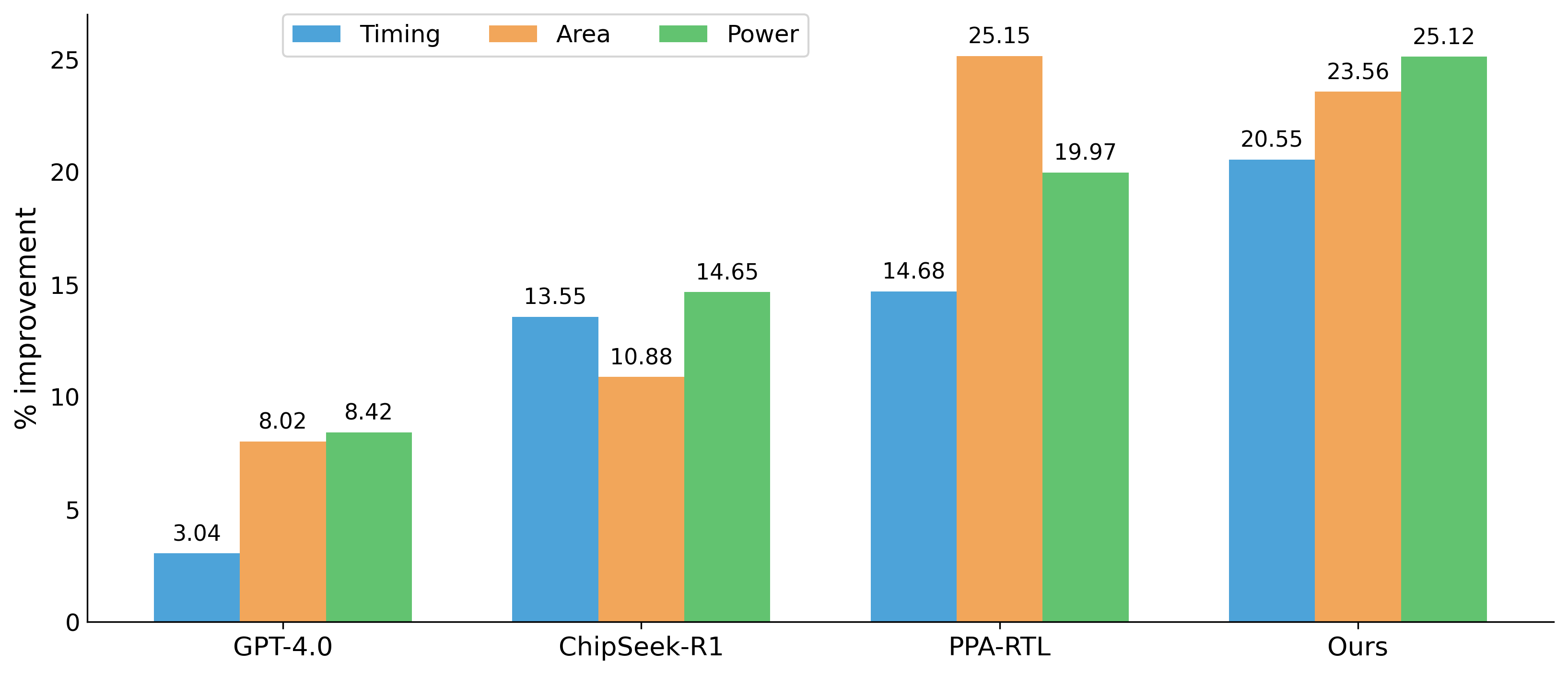}
    \caption{Average \% improvement over RTLLM v2.0.}
    \label{fig:pic4}
\end{figure}

Figure~\ref{fig:pic4} compares our average PPA gains on RTLLM v2.0 against three representative LLM-based baselines reported in prior work. Under the same evaluation protocol (percent improvement relative to the RTLLM v2.0 baseline), a \textbf{pure GPT-4.0 generation} approach yields modest gains (3.04\% timing, 8.02\% area, 8.42\% power). \textbf{ChipSeek-R1} improves over GPT-4.0 (13.55\% timing, 10.88\% area, 14.65\% power) but remains consistently below our approach. \textbf{PPA-RTL} is competitive on area (25.15\%) and achieves solid power reduction (19.97\%), yet our method achieves substantially stronger and more balanced improvements overall, especially on \textbf{timing} and \textbf{power} (20.55\% and 25.12\%), while remaining comparable on area (23.56\%). In terms of utility (Equation~\ref{eq:utility_ud}), our method also achieves the highest overall score (23.08), outperforming PPA-RTL (19.93), ChipSeek-R1 (13.03), and pure GPT-4.0 generation (6.49).These results suggest that reward-model-guided candidate scoring and selection within a synthesis-in-the-loop pipeline is critical for translating LLM-generated RTL edits into consistent, design-wide PPA gains.

\subsection{Ablation Study}

Table~\ref{tab:ablation_budget} summarizes an ablation study under a fixed DC evaluation budget, reporting the maximum normalized best-so-far utility achieved by each variant. Starting from the full method, we remove key components of the pipeline—reward-based selection, candidate dropping/filters, multi-strategy exploration, normalization/deduplication, and diversity-aware selection—to quantify their individual contributions. The resulting utilities show how much each component matters for reaching high-quality PPA tradeoffs within the same DC budget, with the largest degradations indicating the most critical mechanisms.

\begin{table}[htbp]
\centering
\small
\caption{Ablation study under a fixed DC evaluation budget. We report the maximum normalized best-so-far utility achieved (baseline=0).}
\setlength{\tabcolsep}{7pt}
\begin{tabular}{lc}
\toprule
\textbf{Variant} & \textbf{Best utility ($\uparrow$)} \\
\midrule
Full (ours) & 0.98 \\
\midrule
No reward: Random-after-filter & 0.72 \\
No reward: Proxy-only heuristic & 0.79 \\
\midrule
Weaker drop: Compile-only (no sim) & 0.81 \\
No drop: No compile/sim filter & 0.60 \\
\midrule
No exploration: Single-strategy & 0.85 \\
No exploration: Reduced strategy set & 0.88 \\
\midrule
No dedup/normalization & 0.90 \\
No diversity selection & 0.92 \\
\bottomrule
\end{tabular}
\label{tab:ablation_budget}
\end{table}


\section{Conclusion}

This paper presents an LLM-in-the-loop RTL optimization pipeline that makes multi-objective PPA improvement practical under realistic synthesis constraints by combining functional filtering, proxy-guided scoring, preference-trained reward modeling, uncertainty-aware acquisition, and diversity-driven selection to concentrate DC evaluations on high-value candidates. Beyond the selection machinery, our persistent best-tradeoff set $G$ is a core mechanism that ensures we \emph{always improve} in a best-so-far sense: by retaining and updating non-dominated designs, $G$ prevents regression, supports monotonic progress in the discovered set, and provides strong warm-start seeds that enable small, compounding refinements across iterations. Experiments on RTLLM v2.0 across multiple technology libraries demonstrate consistent post-synthesis PPA improvements, while budget-matched ablations verify that reward-guided selection and aggressive candidate dropping drive the majority of sample-efficiency improvements. Overall, maintaining $G$ alongside learning-guided, diversity-aware acquisition turns LLM-generated RTL edits into reliable, cumulative PPA optimization and returns multiple high-quality tradeoff points rather than a single scalar winner.

\bibliographystyle{ieeetr}
\bibliography{ref}

\end{document}